\documentclass[fleqn,twoside]{article}
\usepackage{amsmath}
\usepackage{espcrc2}
\usepackage{cite}

 \newcommand\Refr[1]     {Ref.\,\cite{#1}}
 \newcommand\Refrs[1]    {Refs.\,\cite{#1}}

 \newcommand\Eqn[1]     {Eq.\,(\ref{#1})}
 \newcommand\Eqns[2]    {Eqs.\,(\ref{#1}) and~(\ref{#2})}
 \newcommand\Eqnss[2]   {Eqs.\,(\ref{#1})--(\ref{#2})}

 \newcommand\nt         {\notag}

 \def\beq{\begin{equation}}
 \def\eeq{\end{equation}}
 \def\bsp#1\esp{\begin{split}#1\end{split}}
 \def\bal#1\eal{\begin{align}#1\end{align}}

 \def\beeq{\begin{eqnarray}}
 \def\eeeq{\end{eqnarray}}

 \newcommand\bom[1]     {{\mbox{\boldmath $#1$}}}

 \newcommand\as         {\ensuremath{\alpha_{\mathrm{s}}}}

 \newcommand{\CF}       {C_{\mathrm{F}}}
 \newcommand{\CA}       {C_{\mathrm{A}}}
 \newcommand{\TR}       {T_{\mathrm{R}}}
 
 \newcommand{\Nf}       {\ensuremath{n_{\mathrm{f}}}}
 
 \newcommand{\bT}       {\bom{T}}

 \newcommand\qb         {{\bar q}}

 \newcommand{\ep}       {\epsilon}
 \newcommand\Oe[1]      {\ensuremath{\mathrm O(\ep^{#1})}}

 \newcommand{\rd}       {{\mathrm{d}}}
 \newcommand{\PS}[1]    {\rd\phi_{#1}}
 \newcommand\tsig[1]    {\sigma^{\mathrm{#1}}}
 \newcommand\dsig[1]    {\rd\sigma^{{\rm #1}}}
 \newcommand\dsiga[2]   {\rd\sigma^{{\rm #1,A}_{\scriptscriptstyle #2}}}

 \newcommand\SME[3]     {|{\cal M}_{#1}^{(#2)}{(#3)}|^2}
 \newcommand\M[2]       {\ensuremath{|{\cal{M}}_{#1}^{#2}|^2}}

 \newcommand{\rC}       {{\mathrm C}}
 \newcommand{\rS}       {{\mathrm S}}
 \newcommand{\rSCS}     {{\rC}\kern-2pt{\rS}}

 \newcommand{\IcC}[2]   {{\rC}_{#1}^{#2}}
 \newcommand{\IcS}[2]   {{\rS}_{#1}^{#2}}
 \newcommand{\IcCS}[3]  {{\rC}_{#1}^{~}{\rS}_{#2}^{#3}}
 \newcommand{\IcSCS}[2] {\rC\kern-2pt\rS_{#1}^{#2}}

 \newcommand{\bI}       {\bom{I}}
 \newcommand{\bcA}[1]   {{\bom{\cal A}}_{#1}}

 \newcommand{\ti}[1]    {\tilde{#1}}
 \newcommand{\mom}[1]   {\{p\}^{#1}}
 \newcommand{\momt}[1]   {\{\ti{p}\}^{#1}}

\title{\vspace{-4cm}
\hfill {\normalsize DESY 10-096}\vspace{3.25cm}\\
NNLO jet cross sections by subtraction}

\author{G. Somogyi\address[DESY-Zeuthen]{DESY,\\ Platanenallee 6, 
        D-15738 Zeuthen, Germany}\thanks{Talk presented at 
Loop and Legs in Quantum Field Theory, W\"{o}rlitz, Germany, 
25th-30th April 2010.},
        P. Bolzoni\addressmark[DESY-Zeuthen],
        Z. Tr\'ocs\'anyi\address{CERN PH-TH, on leave from University
of Debrecen and Institute of Nuclear Research of HAS,\\ H-4001 P.O.Box 51, 
Hungary}}
       
\begin{document}

\begin{abstract}
We report on the computation of a class of integrals that appear 
when integrating the so-called iterated singly-unresolved approximate 
cross section of the NNLO subtraction scheme of 
\Refrs{Somogyi:2005xz,Somogyi:2006cz,Somogyi:2006da,Somogyi:2006db}, over the 
factorised phase space of unresolved partons. The integrated approximate 
cross section itself can be written as the product of an insertion 
operator (in colour space) times the Born cross section. We give selected 
results for the insertion operator for processes with two and three 
hard partons in the final state.
\end{abstract}


\maketitle

%
%

\section{Introduction}

High energy particle collisions frequently lead to final states with 
hadronic jets. Jet observables can be ideal for precision studies, since 
their large production cross sections allow them to be measured with 
high statistical accuracy. Examples include: the determination of the 
strong coupling, $\as$, from jet rates and event shapes in electron-positron 
annihilation; the measurement of gluon parton distribution functions 
(and also $\as$) in deep inelastic lepton-hadron scattering into two plus 
one jets; the determination of parton distributions in hadron-hadron 
scattering from single jet inclusive production and vector boson plus jet 
production. The relevant observables are often measured with experimental 
precision of a few per cent or better, thus theoretical predictions with 
the same level of accuracy are necessary. This usually requires the 
computation of next-to-next-to-leading order (NNLO) corrections in 
perturbative QCD.

One of the main bottlenecks of the straightforward application of QCD 
perturbation theory at NNLO is the following. Due to the presence of 
infrared (IR) singularities, the finite higher-order corrections are sums 
of pieces which are separately divergent in $d=4$ spacetime dimensions. 
These IR singularities must be regularised and cancelled before any 
numerical computation may be performed. The most common approach for handling 
IR singularities at next-to-leading order (NLO) accuracy is the subtraction 
method. After regularising all contributions by dimensional regularisation 
in $d=4-2\ep$ dimensions, one builds subtraction terms that simultaneously 
cancel both the kinematical singularities in real-emission phase space 
integrals and the explicit $\ep$-poles in one-loop virtual corrections 
\cite{Catani:1996vz}. 
Since the IR singularity structure of QCD amplitudes is universal, such 
subtraction terms can be defined in a general, i.e.\ process and observable 
independent fashion.

In recent years, severe efforts have been made to extend the subtraction 
method to NNLO accuracy 
\cite{Somogyi:2005xz,Somogyi:2006cz,Somogyi:2006da,Somogyi:2006db,Weinzierl:2003fx,Weinzierl:2003ra,Frixione:2004is,GehrmannDeRidder:2005cm,Daleo:2006xa,Daleo:2009yj,Glover:2010im}, 
and this has proved to be a challenging problem.
In broad terms, any proposed subtraction scheme must address two quite 
distinct difficulties. First, one must define subtraction terms that 
properly regularise the real-emission phase space integrals. In a rigorous 
mathematical sense, the cancellation of kinematical singularities must be 
local, i.e.\ the subtraction terms must have the same pointwise singular 
behaviour in $d$ dimensions, as the real-emission pieces. Second, one must 
combine the integrated form of these counterterms with the virtual 
contributions, cancelling the IR divergences of the loop matrix elements. 
Again, from the point of view of mathematical rigour, this cancellation 
must be local, i.e.\ it must happen pointwise in phase space. Practically, 
the full locality of the subtraction scheme is also important to ensure good 
numerical efficiency of the algorithm. Finally, the construction should be 
universal (process and observable independent), otherwise tedious adaptation 
of the algorithm to every specific problem becomes necessary.

The construction of a general subtraction scheme with fully local counterterms
requires a lot of careful analytical calculations. In order to spare some
work, various approaches have been proposed to NNLO calculations sacrificing
either full locality or generality for easier analytical (but more cumbersome
numerical) treatment. For example, in the antenna subtraction scheme 
\cite{GehrmannDeRidder:2005cm,Daleo:2006xa,Daleo:2009yj,Glover:2010im} 
azimuthal correlations in gluon splitting are not reproduced (and the 
cancellation of $\ep$-poles in the real-virtual contributions is also 
nonlocal). As a consequence, actual numerical 
computation of total rates \cite{GehrmannDeRidder:2007jk,GehrmannDeRidder:2008ug,Weinzierl:2008iv,Weinzierl:2009nz} and event shapes \cite{GehrmannDeRidder:2007bj,GehrmannDeRidder:2007hr,GehrmannDeRidder:2009dp,Weinzierl:2009ms,Weinzierl:2009yz} in $e^+e^-\to 3$ jet 
production in the antenna scheme requires the use of an auxiliary phase 
space slicing. Another option is to develop dedicated subtraction schemes 
that are applicable only to some specific processes, such as the production 
of colourless final states in hadron-hadron collisions \cite{Catani:2007vq,Catani:2009sm,Catani:2010en}. 
Then, one may even propose to dispense with the subtraction method 
altogether, and adopt a strategy such as sector decomposition (see e.g.
\cite{Heinrich:2008si} and references therein).

Nevertheless, it is possible to address all subtleties, and define 
completely local counterterms for real radiation, as done in full detail 
in \Refrs{Somogyi:2005xz,Somogyi:2006cz,Somogyi:2006da,Somogyi:2006db} 
for the case of colourless initial states. (Work towards an 
extension to hadron-initiated processes is presented in 
\Refr{Somogyi:2009ri}.) 
Then, to make this subtraction scheme an effective tool, one must compute the 
integrals of all subtraction terms over the phase spaces of unresolved 
emissions. 
In this contribution, we briefly review the methods used to compute the 
integrated subtraction terms, and report on the evaluation of the integrated 
iterated singly-unresolved cross section, the term labelled $\mathrm{A}_{12}$, 
which appears in \Eqn{eq:sigmaNNLOm}. However, we begin by recalling the 
essentials of the subtraction scheme of 
\Refrs{Somogyi:2005xz,Somogyi:2006cz,Somogyi:2006da,Somogyi:2006db}.

%
%

\section{Subtraction at NNLO}

Consider the NNLO correction to a generic $m$-jet observable,
\bal
\tsig{NNLO} &=
\int_{m+2}\!\dsig{RR}_{m+2} J_{m+2}
+ \int_{m+1}\!\dsig{RV}_{m+1} J_{m+1}
\nt\\&
+ \int_m\!\dsig{VV}_m J_m\,.
\label{eq:sigmaNNLO}
\eal
The three contributions on the right hand side are separately divergent in 
$d=4$ dimensions, but their sum is finite for IR safe observables. To 
obtain the finite NNLO correction, we first continue analytically all 
integrals to $d=4-2\ep$ dimensions and then rewrite \Eqn{eq:sigmaNNLO} as
\bal
\tsig{NNLO} &=
\int_{m+2}\!\dsig{NNLO}_{m+2}
+ \int_{m+1}\!\dsig{NNLO}_{m+1}
\label{eq:sigmaNNLOfin}\\&
+ \int_m\!\dsig{NNLO}_m\,,
\nt
\eal
that is a sum of three finite integrals where the integrands,
\bal
\dsig{NNLO}_{m+2} &=
	\Big\{\dsig{RR}_{m+2} J_{m+2} - \dsiga{RR}{2}_{m+2} J_{m}
\label{eq:sigmaNNLOm+2}\\&\!\!\!\!\!\!\!\!\!\! 
	-\Big[\dsiga{RR}{1}_{m+2} J_{m+1} - \dsiga{RR}{12}_{m+2} J_{m}\Big]
	\Big\}_{\ep=0}\,,
\nt
\\
\dsig{NNLO}_{m+1} &=
	\Big\{\Big[\dsig{RV}_{m+1} + \int_1\dsiga{RR}{1}_{m+2}\Big] J_{m+1}
\label{eq:sigmaNNLOm+1}\\&\!\!\!\!\!\!\!\!\!\! 
	-\Big[\dsiga{RV}{1}_{m+1} + \Big(\int_1\dsiga{RR}{1}_{m+2}\Big)
	\strut^{{\rm A}_{\scriptscriptstyle 1}}
\Big] J_{m} \Big\}_{\ep=0}\,,
\nt
\\
\intertext{and}
\dsig{NNLO}_{m} &=
	\Big\{\dsig{VV}_m + \int_2\Big[\dsiga{RR}{2}_{m+2} - \dsiga{RR}{12}_{m+2}\Big]
\nt\\&\!\!\!\!\!\!\!\!\!\! 
	+\int_1\Big[\dsiga{RV}{1}_{m+1} + \Big(\int_1\dsiga{RR}{1}_{m+2}\Big)
	\strut^{{\rm A}_{\scriptscriptstyle 1}}\Big]\Big\}_{\ep=0} J_{m}\,,
\label{eq:sigmaNNLOm}
\eal
are integrable in four dimensions by construction. 
The approximate cross sections $\dsiga{RR}{2}_{m+2}$ and $\dsiga{RR}{1}_{m+2}$ 
regularise the doubly- and singly-unresolved limits of the real-emission 
piece, $\dsig{RR}_{m+2}$ respectively. The double subtraction due to the \
overlap of these two terms is compensated by $\dsiga{RR}{12}_{m+2}$. 
These terms are given explicitly in \Refr{Somogyi:2006da}. 
Finally, $\dsiga{RV}{1}_{m+1}$ and 
$\Big(\int_1\dsiga{RR}{1}_{m+2}\Big)\strut^{{\rm A}_{\scriptscriptstyle 1}}$ regularise 
the singly-unresolved limits of $\dsig{RV}_{m+1}$ and $\int_1\dsiga{RR}{1}_{m+2}$ 
respectively. They are given explicitly in \Refr{Somogyi:2006db}.

The construction of each approximate cross section in 
\Eqnss{eq:sigmaNNLOm+2}{eq:sigmaNNLOm} is based on the known and universal 
IR limits of tree level and one-loop squared matrix elements, and proceeds 
in two steps. First, the IR factorisation formulae are written in such a 
way that their complicated overlap structure can be disentangled (``matching 
of limits'') \cite{Somogyi:2005xz,Nagy:2007mn}. 
Second, we define ``extensions'' of the formulae, so that 
they are unambiguously defined away from the strict IR limits \cite{Somogyi:2006cz,Somogyi:2006da,Somogyi:2006db}. 
These extensions are defined by the use of various momentum mappings 
that map a set of $m+1$ or $m+2$ momenta into a set of $m$ momenta, 
\beq
\mom{}_{m+1} \longrightarrow \momt{}_{m}
\quad\mbox{and}\quad
\mom{}_{m+2} \longrightarrow \momt{}_{m}\,,
\eeq
such that (i) the delicate structure of cancellations among the matched 
limit formulae in various limits is respected (ii) exact momentum conservation 
is implemented, and (iii) the original $m+1$ or $m+2$ particle phase space 
factorises exactly into the product of an $m$ particle phase space and a 
one- or two-particle phase space measure,
\bal
\PS{m+1}(\mom{}_{m+1};Q) &= \PS{m}(\momt{}_{m};Q) [\rd p_{1,m}]
\\
\intertext{and}
\PS{m+2}(\mom{}_{m+2};Q) &= \PS{m}(\momt{}_{m};Q) [\rd p_{2,m}]\,.
\eal
To finish the definition of the scheme, one must compute once and for all 
the one- and two-particle integrals, denoted formally as $\int_1$ and 
$\int_2$, appearing in \Eqns{eq:sigmaNNLOm+1}{eq:sigmaNNLOm}. 
We discuss this next.

%
%

\section{Integrating the counterterms}

The actual computation of the integrated counterterms leads to a large
number of multi-dimensional integrals. The ultimate goal is to find the
analytical form of the coefficients of a Laurent-expansion (in $\ep$)
of these integrals, which turns out to be a rather a tedious job. In
order to compute these coefficients as efficiently as possible, we have
explored several methods.

First, it is possible to extend the method of integration-by-parts identities 
and solving of differential equations, developed for computing multi-loop 
Feynman integrals \cite{Kotikov:1991pm,Remiddi:1997ny}, 
to the relevant phase space integrations \cite{Aglietti:2008fe}. This method 
yields $\ep$-expansions with fully analytical coefficients, with the final 
results being expressed in terms of two-dimensional harmonic polylogarithms 
(after a suitable basis extension, see \Refr{Aglietti:2008fe} for details).
This approach was used successfully to compute a class of singly-unresolved 
integrals \cite{Aglietti:2008fe}. 

Second, the phase space integrals that arise can be computed via the method 
of Mellin--Barnes (MB) representations \cite{Smirnov:1999gc,Tausk:1999vh,Smirnov:2004ym}. 
Here we obtain the $\ep$-expansion 
coefficients in terms of complex contour integrals over $\Gamma$-functions. 
Performing these integrals by the use of the residue theorem, a 
representation in terms of harmonic sums is obtained. In many cases, the 
sums can be evaluated in a closed form, yielding an analytical result. In 
some instances however, we find multi-dimensional MB integrals that are very 
difficult to compute fully analytically. Nevertheless, in these situations 
a direct numerical evaluation of the appropriate MB representations provides 
a fast and reliable way to obtain final results with small numerical 
uncertainties. We stress that for phenomenological applications, this is 
all that is required, since the relative numerical uncertainty associated 
with phase space integrations is generally much larger than that of the 
integrated counterterms. We have used the MB method to compute all 
singly-unresolved integrals \cite{Bolzoni:2009ye}, 
and very recently to evaluate all 
two-particle integrals appearing in $\int_2 \dsiga{RR}{12}_{m+2}$ as well 
\cite{Bolzoni:2010xx} (see below).

Finally, the method of iterated sector decomposition \cite{Heinrich:2008si} 
can also be used to calculate the integrals we encounter 
\cite{Somogyi:2008fc}. Sector decomposition 
produces a representation of the $\ep$-expansion where the coefficients are
given in terms of (mostly quite cumbersome) finite integrals over the unit 
hypercube. The analytical evaluation of these integrals is not feasible 
except for the simplest cases, and we find that in most instances, the 
MB method provides an integral representation for the expansion coefficients
which is better suited for direct numerical integration. Nevertheless, this 
method is simple to implement and we employed it to numerically cross check 
all our final results. 

%
%

\section{Results}

In previous publications, all one-particle integrals, denoted formally 
by $\int_1$ in \Eqns{eq:sigmaNNLOm+1}{eq:sigmaNNLOm} have been 
evaluated with the methods just discussed \cite{Aglietti:2008fe,Bolzoni:2009ye,Somogyi:2008fc}. 
Here we report on the 
computation of the $\int_2 \dsiga{RR}{12}_{m+2}$ two-particle integrated 
counterterm \cite{Bolzoni:2010xx}. 
To begin, we recall that the iterated singly-unresolved 
subtraction term, $\dsiga{RR}{12}_{m+2}$ can be written symbolically as 
\cite{Somogyi:2006da} 
\beq
\dsiga{RR}{12}_{m+2} = \PS{m} [\rd p_{1,m}] [\rd p_{1,m+1}] 
\bcA{12}\M{m+2}{(0)}\,,
\eeq
where $\bcA{12}\M{m+2}{(0)}$ further evaluates to a sum of several terms, as 
spelt out in detail in \Refr{Somogyi:2006da}. 
Each of these terms is essentially a 
product of kinematical factors times an $m$-parton factorised matrix element, 
the later being independent of the variables in the factorised phase space 
measure $[\rd p_{1,m}] [\rd p_{1,m+1}]$. Thus, it is possible to evaluate the 
integrals of the kinematical factors over the factorised phase spaces, 
without reference to the specific process or observable. The final result, 
after summation over unobserved flavours, can be written in the form of an 
insertion operator (in colour space) times the Born cross section 
\beq
\int_2\dsiga{RR}{12}_{m+2} = \dsig{B}_{m} \otimes 
\bI^{(0)}_{12}(\mom{}_m;\ep)\,.
\eeq
Here the insertion operator has three terms according to the possible colour 
structures
\bal
\bI^{(0)}_{12} &= 
\left[\frac{\as}{2\pi}S_\ep \left(\frac{\mu^2}{Q^2}\right)^\ep\right]^2
\label{eq:I12}\\ & \times
\bigg\{
\sum_{i}
\bigg[\rC^{(0)}_{12,f_i}\,C_{f_i} + \sum_{k\ne i}\rC^{(0)}_{12,f_i f_k}\,C_{f_k}\bigg]
C_{f_i}
\nt\\[2mm]&
+\sum_{j,l\ne j}
\bigg[\rS^{(0),(j,l)}_{12} \CA +\sum_{i} \rSCS^{(0),(j,l)}_{12,f_i} C_{f_i}\bigg]
\bT_j \bT_l
\nt
\eal
\bal
\phantom{\bI^{(0)}_{12}} & 
\nt
+\sum_{i,k\ne i} \sum_{j,l\ne j}
\rS^{(0),(i,k)(j,l)}_{12} \{\bT_i \bT_k , \bT_j \bT_l\}
\bigg\}\,,
\nt
\eal
with $f_i$ denoting flavours, $C_q = \CF \equiv \bT_q^2$,  
$C_g = \CA \equiv \bT_g^2$, and $S_\ep = \frac{(4\pi)^\ep}{\Gamma(1-\ep)}$.
In \Eqn{eq:I12}, the dependence of the functions $\rC^{(0)}_{12,f_i}$, etc.\ 
on the kinematics is suppressed for the sake of simplicity. These functions 
in turn are given as the following combinations of flavour summed integrated 
counterterms
\bal
\rC^{(0)}_{12,f_i} & = 
\label{eq:Ifi}\\[2mm] &\!\!\!\!\!\!\!\!\!\! =
\Big(\IcC{kt}{}\IcC{ktr}{(0)}\Big)_{f_i}
-\Big(\IcC{kt}{}\IcC{ktr}{}\IcSCS{kt;r}{(0)}\Big)_{f_i}
\nt\\[2mm] &\!\!\!\!\!\!\!\!\!\! 
-\Big(\IcC{kt}{}\IcC{rkt}{}\IcS{kt}{(0)}\Big)_{f_i}
+\Big(\IcS{t}{}\IcC{irt}{(0)}\Big)_{f_i}
\nt\\[2mm] &\!\!\!\!\!\!\!\!\!\! 
-\Big(\IcS{t}{}\IcC{irt}{}\IcSCS{ir;t}{(0)}\Big)_{f_i}
-\Big(\IcS{t}{}\IcC{irt}{}\IcS{rt}{(0)}\Big)_{f_i}
\nt\\[2mm] &\!\!\!\!\!\!\!\!\!\! 
+\Big(\IcS{t}{}\IcC{irt}{}\IcSCS{ir;t}{}\IcS{rt}{(0)}\Big)_{f_i}
-\Big(\IcCS{kt}{t}{}\IcC{krt}{(0)}\Big)_{f_i}
\nt\\[2mm] &\!\!\!\!\!\!\!\!\!\! 
+\Big(\IcCS{kt}{t}{}\IcC{krt}{}\IcS{rt}{(0)}\Big)_{f_i}
\nt
\,,
\eal
\bal
\rC^{(0)}_{12,f_if_k} &=
\label{eq:Ififk}\\[2mm] &\!\!\!\!\!\!\!\!\!\! =
\Big(\IcC{kt}{}\IcC{ir;kt}{(0)}\Big)_{f_if_k}
-\Big(\IcC{kt}{}\IcC{ir;kt}{}\IcSCS{kt;r}{(0)}\Big)_{f_if_k}
\nt\\[2mm] &\!\!\!\!\!\!\!\!\!\!
-\Big(\IcCS{kt}{t}{}\IcSCS{ir;t}{(0)}\Big)_{f_if_k}
+\Big(\IcCS{kt}{t}{}\IcSCS{ir;t}{}\IcS{rt}{(0)}\Big)_{f_if_k}
\nt
\,,
\eal
\bal
\rS^{(0),(j,l)}_{12} &=
\label{eq:Ijl}\\[2mm] &\!\!\!\!\!\!\!\!\!\! =
\Big(\IcC{kt}{}\IcS{kt}{(0)}\Big)^{(j,l)}
+\Big(\IcS{t}{}\IcS{rt}{(0)}\Big)^{(j,l)}
\nt\\[2mm] &\!\!\!\!\!\!\!\!\!\!
-\Big(\IcCS{kt}{t}{}\IcS{kt}{(0)}\Big)^{(j,l)}
\nt
\,,
\eal
\bal
\rSCS^{(0),(j,l)}_{12,f_i} &=
\label{eq:Ifijl}\\[2mm] &\!\!\!\!\!\!\!\!\!\! =
\Big(\IcC{kt}{}\IcSCS{kt;r}{(0)}\Big)^{(j,l)}_{f_i}
+\Big(\IcS{t}{}\IcSCS{ir;t}{(0)}\Big)^{(j,l)}_{f_i}
\nt\\[2mm] &\!\!\!\!\!\!\!\!\!\!
-\Big(\IcS{t}{}\IcSCS{ir;t}{}\IcS{rt}{(0)}\Big)^{(j,l)}_{f_i}
-\Big(\IcCS{kt}{t}{}\IcS{rt}{(0)}\Big)^{(j,l)}_{f_i}
\nt
\,,
\eal
\beq
\rS^{(0),(i,k)(j,l)}_{12} =
\Big(\IcS{t}{}\IcS{rt}{(0)}\Big)^{(i,k)(j,l)}\,.
\label{eq:Iikjl}
\eeq
On the right hand sides of these equations, the flavour summed integrated 
counterterms themselves are sums of various integrated subtraction terms. 
Their precise definitions are somewhat long, and will not be reported here. 

Instead, we present results for the insertion operator in \Eqn{eq:I12} for 
processes with two and three partons in the final state. Consider first the 
process $e^+e^-\to 2$ jets. The corresponding Born matrix element is 
$\SME{2}{0}{1_q,2_\qb}$, i.e.\ the quark carries the label 1 and antiquark 
label 2. The colour algebra is trivial \cite{Catani:1996vz}, and all properly 
scaled kinematical invariants are equal to one, i.e.\ $\bI^{(0)}_{12}(\mom{}_2)$ 
does not depend on the kinematics. Introducing the notation \cite{Nagy:1998bb}
$x = \CA/\CF$, $y = \TR/\CF$, we find
\beq
\bsp
\bI^{(0)}_{12}&(\mom{}_{2}) = 
\left[\frac{\as}{2\pi}S_\ep \left(\frac{\mu^2}{Q^2}\right)^\ep\right]^2 
\CF^2
\bigg\{
  \frac{6-2x}{\ep^4}
\\&
+\bigg[12 + \frac{7x}{2} + y\Nf -(4-4x)\Sigma(y_0,D'_0-1)
\\&
-(4-6x)\Sigma(y_0,D'_0)\bigg]\frac{1}{\ep^3} + \Oe{-2}\bigg\}\,,
\esp
\label{eq:I12-2}
\eeq
where $y_0\in (0,1]$ and $D'_0\ge 2$ and integer are fixed parameters 
that enter the precise definition of the subtraction terms. Finally, 
the function $\Sigma(z,N)$ is defined as follows
\beq
\Sigma(z,N) = \ln z - \sum_{k=1}^{N}\frac{1-(1-z)^k}{k}\,.
\eeq
Next, consider $e^+e^-\to 3$ jet production. The Born matrix element is 
$\SME{3}{0}{1_q,2_\qb,3_g}$, i.e.\ the quark carries label 1, the antiquark 
label 2 and the gluon carries label 3. The colour algebra is still 
trivial \cite{Catani:1996vz}, but $\bI^{(0)}_{12}(\mom{}_3)$ now carries genuine 
kinematical dependence. We find
\beeq
&&\!\!\!\!\!\!\!\!\!\!
\bI^{(0)}_{12}(\mom{}_{3})= 
\left[\frac{\as}{2\pi}S_\ep \left(\frac{\mu^2}{Q^2}\right)^\ep\right]^2 
\CF^2
\bigg\{
  \frac{6+2x+x^2}{\ep^4}
\nt\\&&
+\bigg[12 + \frac{101x}{6} +\frac{67x^2}{12}
-\frac{13y}{3}\Nf - \frac{3xy}{2}\Nf
\nt\\&&
-\bigg(4x+\frac{5x^2}{2}\bigg)(\ln y_{13} + \ln y_{23})
\label{eq:I12-3}
\eeeq
\beq
\bsp
\phantom{\bI^{(0)}_{12}}&
\nt
-\bigg(8+x-\frac{5x^2}{2}\bigg)\ln y_{12}
\\&
-(4-4x)\Sigma(y_0,D'_0-1)
\\&
-(4-6x-x^2)\Sigma(y_0,D'_0)
\bigg]\frac{1}{\ep^3} + \Oe{-2}\bigg\}\,.
\esp
\eeq
In this equation $y_{ik}= 2p_i\cdot p_k/Q^2$, with $Q^\mu$ being the total 
incoming momentum. Higher order expansion coefficients (in $\ep$) in 
\Eqn{eq:I12-3} are already quite cumbersome, and will be given elsewhere 
in the form of a computer program \cite{Bolzoni:2010xx}.

%
%

\section{Conclusions}

In this contribution, we have reported the computation of the integrated 
iterated singly-unresolved approximate cross section, 
$\int_2 \dsiga{RR}{12}_{m+2}$, that appears in the NNLO subtraction scheme 
of \Refrs{Somogyi:2006cz,Somogyi:2006da,Somogyi:2006db}. 
The phase space integrals appearing in the computation can 
be evaluated once and for all, and their knowledge (as Laurent expansions 
in $\ep$, up to and including $\Oe{0}$ terms) is necessary to make the 
subtraction scheme an effective tool. All phase space integrals were 
evaluated with two separate methods. First, we used the method of 
Mellin--Barnes representations with harmonic summation to obtain analytical 
results where feasible. In some cases, obtaining complete analytical answers 
is very difficult. In these situations, we integrated the MB representations 
of the expansion coefficients directly, and this provides a fast and reliable 
way to obtain final results with small numerical uncertainties. Second, 
all integrals were computed with the method of sector decomposition as well, 
providing useful numerical checks.

With the evaluation of $\int_2 \dsiga{RR}{12}_{m+2}$, the last task in 
finishing the definition of the subtraction scheme is the computation of 
the integrated doubly-unresolved approximate cross section, i.e.\ 
the term labelled $\mathrm{A}_{2}$ in \Eqn{eq:sigmaNNLOm}. The analytical 
structure and complexity of the integrals that appear in this final
piece are essentially the same as the integrals considered here. 
Therefore, the techniques outlined above will also be applicable to the 
computation of this remaining contribution.

This is work was supported in part by the
Deutsche Forschungsgemeinschaft in SFB/TR 9,
the Helmholtz Gemeinschaft under contract VH-NG-105,
the Hungarian Scientific Research Fund grand OTKA K-60432.

%
%

\end{document}